\documentclass[aps,prl,showpacs,twocolumn,superscriptaddress]{revtex4}
\usepackage{graphicx}
\begin{document}

\title{Systematic tight-binding analysis of ARPES spectra 
of transition-metal oxides}

\author{H.~Wadati}
\email{wadati@phas.ubc.ca}
\homepage{http://www.geocities.jp/qxbqd097/index2.htm}
\affiliation{Department of Physics and Astronomy, University of British
Columbia, Vancouver, British Columbia V6T-1Z1, Canada}

\author{A.~Chikamatsu}
\affiliation{Department of Applied Chemistry, University of Tokyo, 
Bunkyo-ku, Tokyo 113-8656, Japan}

\author{M.~Takizawa}
\affiliation{Department of Physics, University of Tokyo, 
Bunkyo-ku, Tokyo 113-0033, Japan}

\author{H.~Kumigashira}
\affiliation{Department of Applied Chemistry, University of Tokyo, 
Bunkyo-ku, Tokyo 113-8656, Japan}
\affiliation{Core Research for Evolutional Science and 
Technology (CREST), Japan Science and Technology Agency, 
Tokyo 113-8656, Japan}

\author{T.~Yoshida}
\affiliation{Department of Physics, University of Tokyo, 
Bunkyo-ku, Tokyo 113-0033, Japan}

\author{T.~Mizokawa}
\affiliation{Department of Complexity Science and Engineering, 
University of Tokyo, Kashiwa, Chiba 277-8561, Japan}

\author{A.~Fujimori}
\affiliation{Department of Physics, University of Tokyo, 
Bunkyo-ku, Tokyo 113-0033, Japan}

\author{M.~Oshima}
\affiliation{Department of Applied Chemistry, University of Tokyo, 
Bunkyo-ku, Tokyo 113-8656, Japan}
\affiliation{Core Research for Evolutional Science and 
Technology (CREST), Japan Science and Technology Agency, 
Tokyo 113-8656, Japan}

\author{N.~Hamada}
\affiliation{Department of Physics, Tokyo University of Science, 
Chiba 278-8510, Japan}

\date{\today}
\begin{abstract}
We have performed systematic tight-binding (TB) analyses of 
the angle-resolved photoemission spectroscopy (ARPES) 
spectra of transition-metal (TM) oxides A$M$O$_3$ 
($M=$ Ti, V, Mn, and Fe) with the perovskite-type structure 
and compared the obtained 
parameters with those obtained from configuration-interaction (CI) 
cluster-model analyses of photoemission spectra. 
The values of $\epsilon_d-\epsilon_p$ from ARPES are 
found to be similar to the charge-transfer energy $\Delta$ 
from O $2p$ orbitals to empty TM $3d$ orbitals 
and much larger than $\Delta-U/2$ 
($U$: on-site Coulomb energy) expected for 
Mott-Hubbard-type compounds including SrVO$_3$. 
$\epsilon_d-\epsilon_p$ values from {\it ab initio} 
band-structure calculations 
show similar behaviors to those from ARPES. 
The values of the $p-d$ transfer integrals 
to describe the global electronic structure 
are found to be similar in all the estimates, 
whereas additional narrowing beyond the TB description 
occurs in the ARPES spectra of the $d$ band. 
\end{abstract}
\pacs{79.60.-i, 71.15.Mb, 71.28.+d, 71.30.+h}
\maketitle

$3d$ transition-metal (TM) oxides have attracted 
a lot of interest in these decades because of 
their intriguing physical properties such as metal-insulator transition, 
colossal magnetoresistance, and the ordering of 
spin, charge, and orbitals \cite{rev}. 
Photoemission spectroscopy has greatly contributed to the 
understanding of the electronic structures of these materials. 
The local electronic structure of strongly correlated $3d$ 
states of the 
TM atoms hybridizing with ligand $p$ orbitals has 
been understood in the framework of 
the Zaanen-Sawatzky-Allen (ZSA) diagram \cite{ZSA} 
based on the configuration-interaction (CI) 
cluster-model analyses of core-level and valence-band 
angle-integrated photoemission 
spectra \cite{Bocquet2,Saitoh2,cluster1}. 
On the other hand, 
how the energy band structure of the periodic lattice 
emerges from the correlated local electronic structure has  
been an extremely difficult problem as 
has been extensively discussed in the case of 
$f$-electron systems \cite{palee,SIF}, and 
no clear scenario has been established so far. 
On the experimental side, rather surprisingly 
Sarma {\it et al.} \cite{Sarma1} have found that 
the angle-integrated valence-band photoemission spectra 
of various perovskite-type $3d$ TM oxides 
can be well reproduced by band-structure 
calculation based on the local spin-density approximation 
in spite of the strong electron correlation. 

In order to address the issue of a proper description of 
the periodic systems with strong correlation, 
the experimental determination of band dispersions 
by angle-resolved photoemission spectroscopy (ARPES) 
has been highly required, but such data have been largely 
limited to low dimensional compounds 
\cite{hightcrmp, MnARPESNature, MnARPESNature2} 
of limited chemical variety. 
For three-dimensional perovskite oxides, which 
have a wide variety of chemical compositions, 
even when bulk single crystals are available, 
they do not have 
cleavage planes required for ARPES measurements. 
Recently, technological development has made it possible to 
grow high-quality single-crystal thin films 
using the pulsed laser deposition method, 
and a setup has been developed for their film growth 
followed by {\it in-situ} photoemission 
measurements \cite{Horiba, LSCOfilm, Shi, Shi2}. 
Since ARPES data have accumulated for various $3d$ TM oxides of 
perovskite structures 
[SrTiO$_3$ (STO) \cite{MaekawaSTO}, 
SrVO$_3$ (SVO) \cite{TakizawaSVO}, 
La$_{1-x}$Sr$_x$MnO$_3$ (LSMO) \cite{Chika, Chika2}, 
Pr$_{1-x}$Ca$_x$MnO$_3$ (PCMO) \cite{WadatiPCMOPRL} and 
La$_{1-x}$Sr$_x$FeO$_3$ (LSFO) \cite{LSFOARPES}] 
and have been analyzed 
using a parameterized nearest-neighbor 
tight-binding (TB) model \cite{Kahn,ReO3,STO,Wadatiphase}, 
we are in a position to look at the systematic variation 
of the electronic structure of the periodic lattice 
of $3d$ TM atoms and derive empirical rules, as has been done 
for the local electronic structure by ZSA \cite{ZSA}. 
In particular, since there is no established theoretical 
description for correlated periodic systems, 
it is highly desirable to obtain relationship 
between parameters in CI theory and those 
describing ARPES band structures. 
In this paper, we examine the TB parameters 
that describe the ARPES band dispersion 
in order to deduce systematic variations of 
the energy level ($\epsilon_d$) 
of TM $3d$ orbitals relative 
to that ($\epsilon_p$) of O $2p$ orbitals and 
the $p-d$ transfer integrals 
[Slater-Koster parameter $(pd\sigma)$] 
and compare 
them with those derived from the 
CI cluster-model analyses of angle-integrated 
photoemission spectra, namely, $(pd\sigma)$, 
the charge-transfer (CT) energy 
from the O $2p$ orbitals to the empty TM $3d$ orbitals 
denoted by $\Delta$, and 
the average $3d-3d$ on-site Coulomb interaction energy 
denoted by $U$. 
For comparison, we have also fitted 
the 
{\it ab initio} band structure 
to the same TB model, as has been done 
in Ref.~\cite{sarmaLDATB}. 
The values of $\epsilon_d-\epsilon_p$ 
from ARPES were found to be similar to $\Delta$ 
rather than $\Delta-U/2$ expected for 
Mott-Hubbard (MH) type compounds. 
The values of $|(pd\sigma)|$ 
(determined from the O $2p$ band dispersions) 
were found to be similar in these estimates, but 
additional narrowing in the $d$ band 
(by a factor of $0.3-0.5$) 
was seen in ARPES beyond the TB description. 

All the ARPES data analyzed in this paper 
were measured using 
a photoemission spectroscopy system combined 
with a laser molecular beam epitaxy chamber 
at beamline BL-1C of the Photon Factory, 
KEK \cite{Horiba}. Details of the experimental conditions 
for the respective materials were described in 
Refs.~\cite{MaekawaSTO, TakizawaSVO, Chika, Chika2, 
WadatiPCMOPRL, LSFOARPES}. 
The {\it ab initio} band-structure calculations were performed 
using the linearized augmented plane wave method 
implemented in the WIEN2K package \cite{wien}. 
The exchange and correlation effects are treated
within the generalized gradient 
approximation \cite{gga}. 
We assumed a cubic paramagnetic (PM) state, 
and the lattice constants were taken as 
$3.91 \mbox{\AA}$ for STO, 
$3.84 \mbox{\AA}$ for SVO, 
$4.04 \mbox{\AA}$ for LaMnO$_3$ (LMO), and 
$4.02 \mbox{\AA}$ for LaFeO$_3$ (LFO). 
We fitted the ARPES and {\it ab initio} results 
to parameterized nearest-neighbor 
TB band structures 
\cite{Kahn,ReO3,STO,Wadatiphase}. 
We assumed the PM state 
for the ARPES results of STO and SVO, 
the ferromagnetic (FM) state for LSMO, 
the C-type antiferromagnetic (AF) state for PCMO, 
and the G-type AF state for LSFO. 
The size of the matrix for the calculations were 
$14\times 14$ (TM $3d$ orbitals: 5, 
oxygen $2p$ orbitals: $3\times 3 =9$) in the cases of 
PM and FM states, and $28\times 28$ in the cases of 
AF states. 
The effects of ferromagnetism and antiferromagnetism 
were treated phenomenologically by assuming an energy difference 
$\Delta E$ between the spin-up and spin-down TM sites. 
Parameters to be adjusted are $\Delta E$, 
the energy level of the TM $3d$ orbitals 
relative to that of the O $2p$ orbitals, $\epsilon_d-\epsilon_p$, 
and Slater-Koster parameters 
$(pd\sigma)$, $(pd\pi)$, $(pp\sigma)$, 
and $(pp\pi)$ \cite{JC}. In the cases of the FM and 
AF states, $\epsilon_d$ is the average of 
the up-spin $d$ level $\epsilon_d-\Delta E/2$ and 
the down-spin $d$ level $\epsilon_d+\Delta E/2$. 
The details of the matrix elements for the 
calculations were described in Refs.~\cite{ReO3, Wadatiphase}. 

In the CI cluster-model 
calculations \cite{Bocquet2,Saitoh2,cluster1,earlyCI}, 
a TMO$_6$ octahedral cluster is considered. 
In this model, the ground state is described as 
$$
\Psi_g=\alpha_1|d^n\rangle+\alpha_2|d^{n+1}\underline{L}\rangle+
\alpha_3|d^{n+2}\underline{L}^2\rangle+\cdots.
$$
The final state for the emission of a TM $2p$ core electron is given by 
$$
\Psi_f=\beta_1|\underline{c}d^n\rangle+\beta_2|\underline{c}d^{n+1}\underline{L}\rangle+
\beta_3|\underline{c}d^{n+2}\underline{L}^2\rangle+\cdots,
$$
where $\underline{c}$ denotes a TM $2p$ core hole. 
The final state for the emission of a TM $3d$ electron is given by
$$
\Psi_f=\gamma_1|d^{n-1}\rangle+\gamma_2|d^n\underline{L}\rangle+
\gamma_3|d^{n+1}\underline{L}^2\rangle+\cdots. 
$$
In this work, we have taken 
the parameter values of STO and SVO from 
Ref.~\cite{earlyCI}, and those for LSMO and LSFO 
from Refs.~\cite{horibaLSMO,Wadati}, where 
these values have been obtained by 
fitting the TM $2p$ 
core-level and valence-band photoemission 
spectra. 
 
\begin{figure}
\begin{center}
\includegraphics[width=9cm]{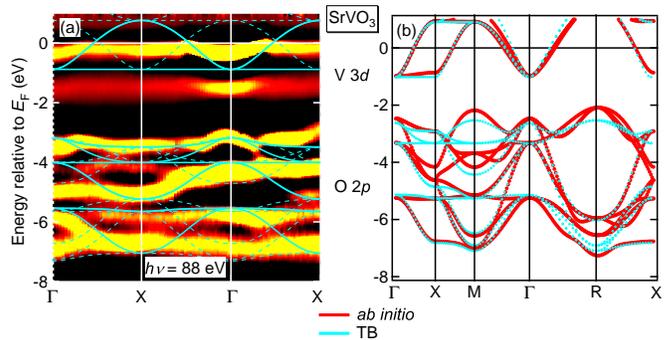}
\caption{(Color online) 
Tight-binding fit to 
the ARPES spectra \cite{TakizawaSVO} (a) and 
{\it ab initio} band-structure calculation (b) for SrVO$_3$. 
In panel (a), bright parts 
correspond to energy bands.}
\label{fig1}
\end{center}
\end{figure}

In Fig.~\ref{fig1}, we show an 
example of the TB fit to 
the ARPES spectra and 
{\it ab initio} band-structure calculation. 
Figure \ref{fig1} (a) shows  
a fit to the ARPES spectra of 
SVO \cite{TakizawaSVO}. 
The bands below $\sim -3$ eV are mainly from O $2p$ states, 
and those between $\sim -2$ eV and $E_F$ are mainly from V $3d$ states. 
The dispersions of the O $2p$ bands were well 
reproduced by the calculation, but 
the widths of the V $3d$ bands were about 
50\% of the calculated ones. 
Such a mass renormalization is also observed 
in the ARPES spectra of LSMO \cite{Chika} and 
high-$T_C$ cuprates \cite{hightcrmp}. 
Figure \ref{fig1} (b) shows 
a TB fit to the {\it ab initio} band-structure calculation 
of SVO. The parameters 
$\epsilon_d - \epsilon_p = 3.6$ eV and 
$(pd\sigma) = -2.2$ eV give a fairly good fit to 
the {\it ab initio} calculation. 

\begin{figure}
\begin{center}
\includegraphics[width=9cm]{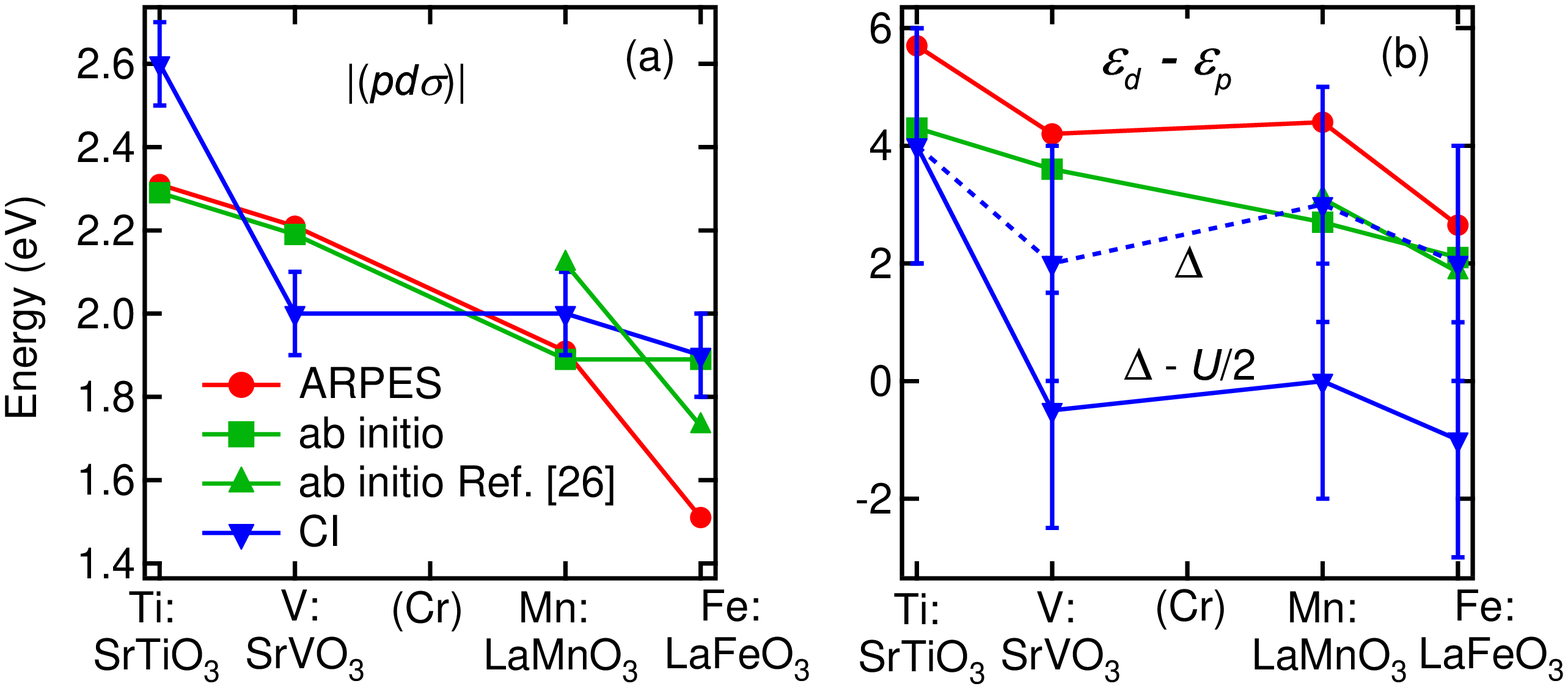}
\caption{(Color online) 
Material dependences of 
the $p-d$ transfer integral ($|(pd\sigma)|$) (a) and 
the energy level of TM $d$ 
relative to O $2p$ ($\epsilon_d-\epsilon_p$) (b) deduced from 
ARPES and {\it ab initio} calculations compared with those from 
CI calculation. 
Values from {\it ab initio} calculations 
in Ref.~\cite{sarmaLDATB} has been also plotted. 
The CI parameter values of STO and SVO are taken from 
Ref.~\cite{earlyCI}, and those for LSMO and LSFO are 
from Refs.~\cite{horibaLSMO,Wadati}.}
\label{fig2}
\end{center}
\end{figure}

Figure \ref{fig2} shows comparison of the variation 
of $|(pd\sigma)|$ [panel (a)] and $\epsilon_d-\epsilon_p$ [panel (b)] 
for ARPES and {\it ab initio} compared 
with parameter values deduced from 
CI cluster-model analysis. 
As seen from Fig.~\ref{fig2} (a), the values of 
$(pd\sigma)$ are similar in all the three estimates. 
This means that, as far as 
the strength of $p-d$ hybridization is concerned, 
the values from the local CI cluster-model analysis 
can be used to describe the band structures of periodic systems. 
The behavior of $\epsilon_d-\epsilon_p$ from ARPES 
follows $\Delta$ from CI analysis, 
as shown in Fig.~\ref{fig2} (b). 
The $\epsilon_d-\epsilon_p$ values from {\it ab initio} calculations 
are similar to or a little (by $\sim 1$ eV) smaller than 
those from ARPES. 
This reflects the well-known tendency of {\it ab initio} calculations 
to underestimate the band gaps and 
the binding energies of the O $2p$ bands. 
One typical example is the underestimation of the 
values of the bad gaps in STO \cite{JAPSTO}. 
The indirect ($R\rightarrow \Gamma$) and 
direct ($\Gamma \rightarrow \Gamma$) gaps are 
calculated to be 1.9 eV and 2.2 eV, while 
the experimental values are 3.25 eV and 3.75 eV. 
However, as this difference is rather 
small, one may conclude that 
{\it ab initio} calculations can describe the 
angle-integrated photoemission spectra of these materials 
fairly well \cite{Sarma1}. 
It is remarkable to see that, except for the common shift of 
the O $2p$ band, the experimental $\epsilon_d-\epsilon_p$ of 
complicated origin 
is well reproduced by the simple 
{\it ab initio} calculation. 

For the $d^0$ compound (STO), 
the value of $\epsilon_d-\epsilon_p$ is 
expected to be given by $\Delta$. 
In the case of MH-type compounds 
such as SVO, $\epsilon_d-\epsilon_p$ is expected to be given by 
$\Delta-U/2$ \cite{sarmaLDATB}. 
However, Fig.~\ref{fig2} shows that 
the $\epsilon_d-\epsilon_p$ values from ARPES are closer 
to $\Delta$ than $\Delta-U/2$ for all compounds 
including SVO. 
One should note here that 
the values 
of $\Delta$ shows strong dependence on TM atoms and changes 
from $\sim 0$ to $\sim 6$ eV, 
whereas those of $U$ do not strongly depend on TM atoms and 
are as large as $5 - 8$ eV \cite{Bocquet2, earlyCI}. 

\begin{figure}
\begin{center}
\includegraphics[width=8cm]{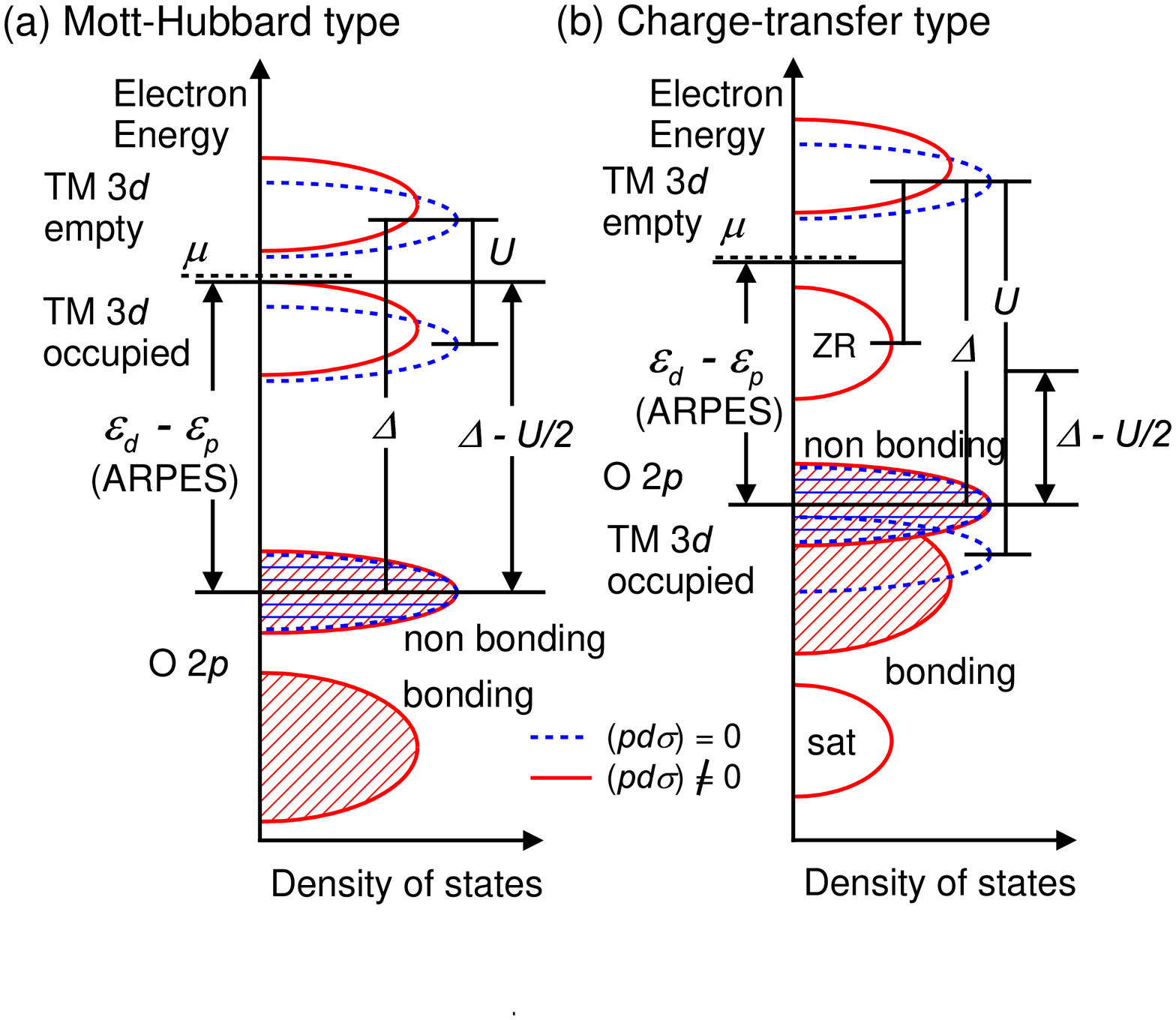}
\caption{(Color online) Energy diagrams for 
Mott-Hubbard-type (a) and 
charge-transfer-type (b) compounds. 
Sat: charge-transfer satellite ($d^n$); 
ZR: Zhang-Rice singlet-like states ($d^n\underline{L}$). Here, 
$\epsilon_d-\epsilon_p$ is the $3d$ level relative 
to O $2p$, $\Delta$ is the charge-transfer energy 
from O $2p$ to the empty TM $3d$, and 
$U$ is the $3d-3d$ on-site Coulomb interaction energy.}
\label{fig3}
\end{center}
\end{figure}

To understand the difference in $\epsilon_d-\epsilon_p$ 
between ARPES and the simple MH picture even for SVO, 
we consider the single-particle energy diagrams 
for MH type compounds, 
where $\Delta>U$, [panel (a)] 
and CT type compounds, 
where $\Delta<U$, [panel (b)] 
and the assignments of 
band characters as shown in Fig.~\ref{fig3}. 
In a MH-type compound, 
the TM $3d$ states are located 
above the O $2p$ band, and 
the $p-d$ hybridization simply causes 
upward and downward shifts of the $p$ and $d$ levels, 
respectively, leading to little difference between 
ARPES and CI. 
Therefore, in pure MH-type compounds, 
the relationship 
$\epsilon_d-\epsilon_p=\Delta-U/2$ holds, 
which is inconsistent with our results. 
In a CT-type compound, the effect of 
$p-d$ hybridization is complicated, creating new states 
in the occupied part, i.e., in the electron removal spectrum. 
That is, a charge-transfer satellite of $d^{n-1}$ 
final-state character appears below the O $2p$ band 
and the split-off state of 
the Zhang-Rice (ZR) states ($d^n\underline{L}$ final-state character) 
are formed above the O $2p$ band as a result of $p-d$ 
hybridization, as shown in Fig.~\ref{fig3} (b). 
In the case of a $d^9$ system (Cu$^{2+}$), 
the ZR state becomes a spin singlet, well known as the  
ZR singlet \cite{ZRS}. 
The emergence of the charge-transfer 
satellite and the ZR state can 
be understood from the Anderson impurity model, where 
the discrete TM $3d$ impurity states interact with the 
O $2p$ continuum states through strong $p-d$ hybridization 
and two new split-off states are created. 
When we fit the ARPES results, 
we consider the ZR state as ``effective'' TM $3d$ bands. 
This explains why 
the $\epsilon_d-\epsilon_p$ values in ARPES becomes considerably 
larger than $\Delta-U/2$ in the CT-type compounds 
as shown in Fig.~\ref{fig3}(b). 
The difference of these two is large also in SVO, 
which has long 
been considered as a typical MH compound. 
This result shows that SVO is not a pure MH compound 
and hybridization between O $2p$ and V $3d$ states is 
comparable to CT-type compounds, as already 
suggested in Fig.~15 in Ref.~\cite{earlyCI} and 
concluded by the recent 
CI calculation in Ref.~\cite{SVOCI}. 

The similarity of the $\epsilon_d-\epsilon_p$ values in 
{\it ab initio} and ARPES shows 
that in {\it ab initio} $\epsilon_d-\epsilon_p$ is similar to $\Delta$ 
not to $\Delta-U/2$. 
This result means that the $d$ bands in {\it ab initio} calculations 
are not the pure $d$ bands before $p-d$ hybridization 
but the ZR state after being 
considerably hybridized with O $2p$ states. 

Finally, we would like to comment the limitation of our 
analyses of the ARPES spectra. 
The values of $|(pd\sigma)|$, which describes 
the global electronic structure, were similar 
in the cases of ARPES, {\it ab initio}, and CI, but 
the $d$-band narrowing observed in ARPES were 
about $1/2$ or $1/3$, which cannot be 
described in the TB band-structure calculation. 
This is the effect of electron correlation 
at the TM sites. 

In conclusion, we have compared the parameters obtained from 
TB fit to the ARPES spectra of TM oxides 
A$M$O$_3$ ($M=$ Ti, V, Mn, and Fe) and 
those obtained by TB fit to the {\it ab initio} 
band-structure calculation 
with those from CI cluster-model calculations. 
The $\epsilon_d-\epsilon_p$ values obtained 
by TB fit to {\it ab initio} band-structure 
calculations 
were a little smaller than those in ARPES, reflecting 
the general tendency of {\it ab initio} calculations 
to underestimate the binding 
energies of O $2p$ bands. 
The $\epsilon_d-\epsilon_p$ values from ARPES 
were significantly larger than $\Delta-U/2$ 
for all the compunds including SVO, which has been 
considered as a typical MH system. 
We attribute this trend to the 
ZR-band origin of the effective $d$ band. 
The values of $|(pd\sigma)|$ were found to be similar in all 
the three estimates, showing the validity of using parameters 
obtained from the local cluster or impurity Anderson models 
to describe the global electronic 
band structure of the TM oxides. 
However, the narrowing of the (``{\it effective}'') $d$ 
band seen in ARPES was beyond the TB description. 

The authors would like to thank I. Elfimov and 
G. A. Sawatzky for informative 
discussions. This work was supported by a Grant-in-Aid 
for Scientific Research (A19204037) from 
the Japan Society for the Promotion of 
Science (JSPS) and a Grant-in-Aid 
for Scientific Research in Priority Areas 
``Invention of Anomalous Quantum Materials'' 
from the Ministry of Education, Culture, 
Sports, Science and Technology. 
This work was done under the approval of the 
Photon Factory Program Advisory Committee 
(Proposal Nos.~2002S2-002 and 2005S2-002). 
H. W. acknowledges financial support from JSPS. 

\bibliography{LSFOtex}
\end{document}